\documentclass[review,12p]{elsarticle}

\usepackage{graphicx}
\usepackage{physics}
\usepackage{amssymb}
\usepackage{tabularx,booktabs}
\usepackage{multicol}
\journal{Optik}
\usepackage{amsmath}
\usepackage[super]{nth}
\usepackage[justification=centering]{caption}
\usepackage{scalerel,stackengine}
\usepackage{xcolor}
\usepackage{cleveref}
\usepackage{placeins}
\usepackage{hyphenat}
\usepackage{footnote}

\begin{document}
\begin{frontmatter}

\title{Gated InGaAs Detector Characterization with Sub-Picosecond Weak Coherent Pulses }
\author[1]{Gautam Shaw\corref{cor1}}
\ead{ee15d047@ee.iitm.ac.in}
\cortext[cor1]{Corresponding author}
\author[1]{Shyam Sridharan}
\author[1]{and Anil Prabhakar}

\address[1]{Department of Electrical Engineering, IIT Madras,  Chennai, 600036, India}
\begin{abstract}
We demonstrate  a method to characterize a gated InGaAs single-photon detector (SPD). Ultrashort weak coherent pulses, from a mode-locked sub-picosecond pulsed laser, were used to measure photon counts, at varying arrival times relative to the start of the SPD gate voltage. The difference of detection probabilities for two  gate windows with different widths, and the uneven detection probabilities within a gate window, were used to estimate the afterpulse probability with respect to the detector parameters: excess bias, width of gate window and hold-off time.  We estimated an afterpulse decay time of $1.1$ to $2.4\,\mu$s using a power-law fit to the decay in afterpulse probability. Finally, we  measure the timing jitter of the SPD, as 240~ps, using a time to digital converter with a resolution of 55~ps.
\end{abstract}
\end{frontmatter}

\section{Introduction}
Single-photon detectors  (SPDs) are integral to quantum communication  and quantum information processing systems \cite{zhang2015advances, hadfield2009single}. They  are also widely used in applications such as LIDARS, ultra-fast optical measurements, time-domain reflectometry, and fluorescence lifetime imaging microscopy \cite{diamanti20061, perenzoni2014depth}. In the telecommunication band (C and L band), superconducting nanowire SPDs (SNSPDs), and InGaAs/InP SPDs are the two major choices for single-photon detection. InGaAs based SPDs have a lower detection efficiency and higher dark counts, but remains the favored choice for practical  quantum key distribution technology  due to their ease of operation,  low power consumption, compact construction, and low cost \cite{liu2020reducing, wang2019design}. 

An SPD is operated in a free-running, or gated mode, a choice made based on the specific application. In the free-running mode, a reverse bias voltage is applied periodically above the breakdown voltage of the SPD, and the electric field becomes high enough for a single photon to  trigger an avalanche of charge carriers \cite{mcintyre1999new, lunghi2012advantages}. In the  gated mode, a short  voltage pulse is combined with a reverse bias voltage. The SPD  bias is kept below the breakdown voltage, except when we expect the single photon to arrive. Typically, the gate detection window, or gate widths, are a few nanoseconds. This yields photon detection with reasonable low dark count rate and low afterpulse probability.

Gated mode InGaAs detectors have been thoroughly investigated and experimentally characterized by many  researchers \cite{ribordy1998performance,prochazka2001peltier,hu2011characterization,tosi2012ingaas}. 
One of the drawbacks of using a high excess bias voltage is the afterpulse effect that arises from the release of trapped carriers during successive gate pulses~\cite{kang2003dark}. These carriers could have been trapped in the multiplication layer of SPD during any of the previous avalanche pulses. The release rate of trapped carriers from the multiplication region is proportional to the number of filled traps \cite{hu2011characterization}. The afterpulse probability is related to the lifetime of trapped carriers. It also depends on the hold-off time after a successful detection, size of gate window, temperature, and excess  bias voltage. This effect is particularly troubling in QKD experiments, since they increase the quantum bit error rate (QBER).

Calibrated laser and correlated photon pair method was used to estimate the key parameters: detection efficiency, timing jitter, and dark count probability \cite{hadfield2009single}. Humer et al. demonstrated SPD characterization based on photon counting statistics, whereas a characterization method based on the statistics of times between consecutive photon detections was reported by Da Silva et al. \cite{humer2015simple} \cite{da2011real}. Yen et al described a  technique to measure afterpulsing probability based on the difference of detection efficiency at two different gate window widths~\cite{yen2008simple}.

In this letter, we propose a method to characterize the gated mode InGaAs SPD using a sub-picosecond laser pulse from a mode-locked fiber laser centered at 1563 nm, with a spectral width of 4.7~nm~\cite{Shaw17,Shaw21}. The characterization technique is based on two factors: (1) the difference of detection clicks at two different gate widths and (2) the dependency of afterpulse noise on the arrival time of photon within the gate width of SPD due to  non-uniformity of detection efficiency. We then use a power-law to estimate the half-life time of afterpulse decay for a gated InGaAs SPD.  We also estimate the timing jitter of the SPD  using a time to digital converter (TDC) with a resolution of 55~ps. 

\section{Characterization method}

We have quantified the decaying rate of afterpulse carriers and dark carriers in the SPD separately.  Our measurement method is based on two phenomena: (1) the difference of detection clicks at two different gate window widths \cite{yen2008simple} determines the afterpulse noise, (2) the dependency of afterpulse noise on the arrival time of photon due to  non-uniformity of detection efficiency within the gate width of SPD. To better understand the SPD characterization system, we have listed down the parameters in Table~\ref{table:SPD_char}.

\begin{table}[tbh]
\caption{Terminology used for our experiment} 
\centering 
\fontsize{9}{11}\selectfont
\begin{tabular}{p{2.0cm} p{5.5cm}} 
\hline 
Parameters & Description  \\
\hline 
$T_{\text{laser}}$  & Repetition rate of MLL pulses \\
$T_{\text{rep}}$ & $T_{\text{on}}$+$T_{\text{off}}$ (gate repetition rate) \\
$T_{\text{dark}}$ & Mean time between avalanche events without any incident photons\\
$T_{\text{ph}}$ & Mean time between avalanche events when photons are incident \\
$T_{\text{hold}}$ & Time when detector is quenched \\
$V_{\text{ex}}$& Excess  bias voltage\\
\hline 
\end{tabular}
\label{table:SPD_char} 
\end{table}

\begin{itemize}
    \item $T_{\text{dark}}$ will be in order of milliseconds, since the dark count rate is in order of 100's of counts/s.
    \item $T_{\text{ph}}$ $\geq$ $T_{\text{hold}}$ since we have $T_{\text{rep}}$ $\ll$ $T_{\text{hold}}$, which implies that $T_{\text{ph}}$ $\approx$ $T_{\text{hold}}$.
    \item $T_{\text{laser}}$ = $m.T_{\text{rep}}$, where $m$ is an integer, as we synchronize our FPGA clock to a MLL.
    \item  We have $\frac{T_{\text{ph}}}{T_{\text{laser}}}$=$\eta_\text{det}$.$\mu$ $\approx$ 0.01, where $\eta_\text{det}$ is the detector efficiency and $\mu$ is the mean photon number per pulse.
    \item Excess bias voltage is represented as $V_{\text{ex}}$, and we have  $V_{\text{ex}}$= $V_{\text{r}} - V_{\text{bd}}$, where $V_{\text{r}}$ and $V_{\text{bd}}$ are reverse bias voltage and break-down voltage respectively.

\end{itemize}
\begin{figure}[tbh]	
	\centering
	\includegraphics[width=0.68\columnwidth]{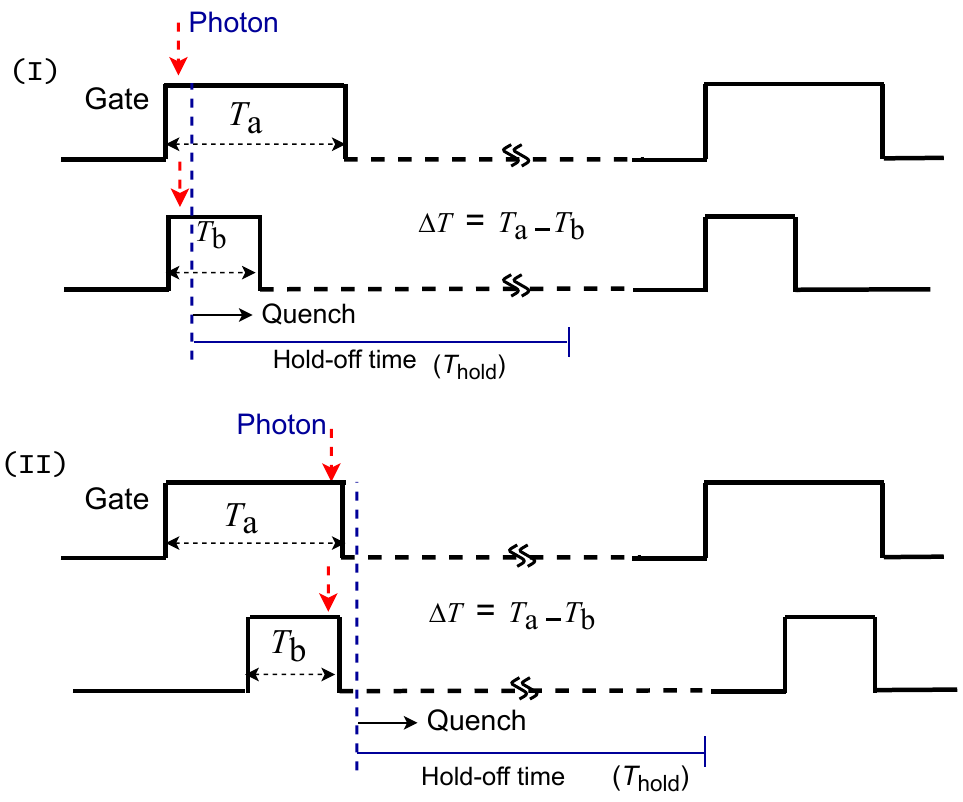}
	\centering
	\caption{Arrival time of photon within the gate ON time is synchronized to cases I: immediately after the rising edge and II: just before the falling edge, of the gate. The gate width  $T_{\text{on}}$ is chosen to be $T_{\text{a}}$ or $T_{\text{b}}$.}
	\label{method}
\end{figure}
Consider the two cases in Fig. \ref{method}: (I) the photon is incident just after the rising edge of the gate, and (II) photon is incident just before the falling edge of gate. Carrier build-up takes place till quenching starts, and a small fraction of carriers remain trapped within the intrinsic layer of the semiconductor. 
Generally, the avalanche current reaches a steady state within a build-up time of a few hundred picoseconds. However, some additional carriers are also generated in this process and trapped in the active region of the SPD. We expect that the density of trapped carriers in $T_{\text{a}}$, with a longer gate ON time, is higher than that of $T_{\text{b}}$. Consequently, we will observe that the afterpulse probability due to an additional gate ON time of $\Delta T$ is proportional to the difference in number of SPD  clicks in Case I and Case II (gate ON times of $T_{\text{a}}$ and $T_{\text{b}}$),  where $\Delta T$ = $T_{\text{a}}$ - $T_{\text{b}}$. 

Referring to Fig. \ref{method}, cases (I) and (II), the afterpulse probability (APP) can be defined as ~\cite{yen2008simple}:
\begin{equation}
\text{APP}(\Delta T)= \frac{{C_{\text{a}}-C_{\text{b}}}}{C_{\text{a}}},
\label{final equation for afterpulse}
\end{equation}
where $C_{\text{a}}$ and $C_{\text{b}}$ are the number of clicks for two different gate ON times, $T_{\text{a}}$  and $T_{\text{b}}$ respectively. The difference in number of clicks ($C_{\text{a}}$ - $C_{\text{b}}$) should be dependent on the values of SPD paramters, i.e, excess-bias voltage, temperature and hold-off time. By varying the value of any one of the SPD parameters, while keeping other parameters fixed, we are able to measure the afterpulse probability for a gate width of time $\Delta T$. 

\section{Experiment set-up}
\begin{figure}[tbh]	
 	\centering
 	\includegraphics[width=0.73\columnwidth]{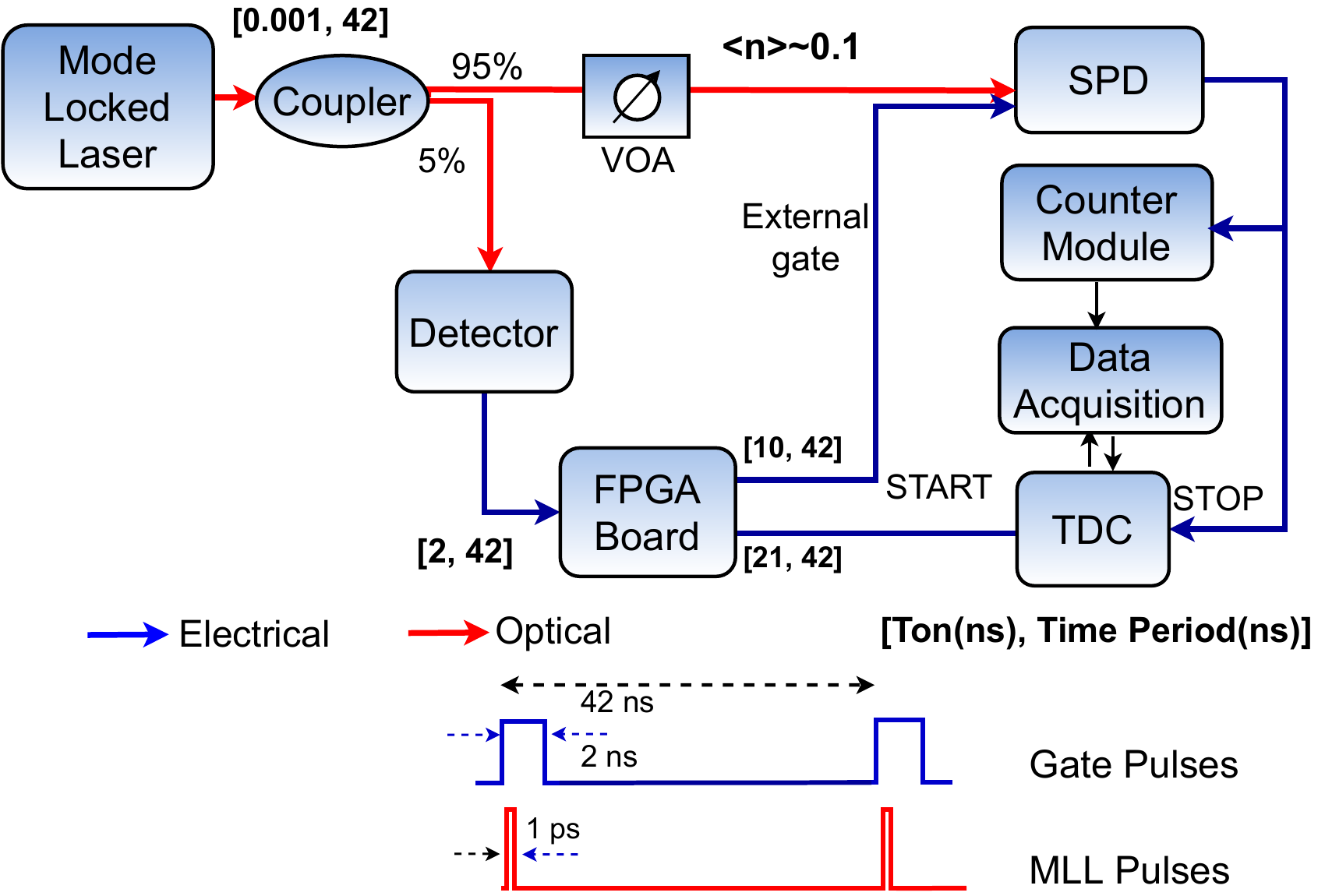}
 	\centering
 	\caption{Afterpulse characterization set-up using a mode-locked fiber laser: Time to digital converter (TDC) is used to measure timing jitter of SPD }
 	\label{fig:afterpulse_setup}
\end{figure}

The schematic of the characterization set-up is shown in the Fig.~\ref{fig:afterpulse_setup}. Short coherent  pulses, generated from a 1563.64~nm mode-locked fiber laser with a spectral width of 4.7~nm \cite{Shaw17,Shaw21}, were attenuated to a mean photon number $\mu \sim 0.1$, using a calibrated variable optical attenuator (VOA). The detector is a gated InGaAs SPD (Micro Photon Devices, Model No. IR-DH-025-C-F). A small fraction $(5\%)$ of the laser pulse was sent to a 5~GHz InGaAs photodetector using an optical splitter, and used to trigger a  field programmable gate array (FPGA) board  synchronous with the mode-locked laser (MLL) pulses.  The FPGA board was programmed with electronic delays  and used to adjust the time delay of gate pulses relative to the arrival time of a photon at the SPD. 

\begin{figure}[tbh]	
 	\centering
 	\includegraphics[width=0.72\columnwidth]{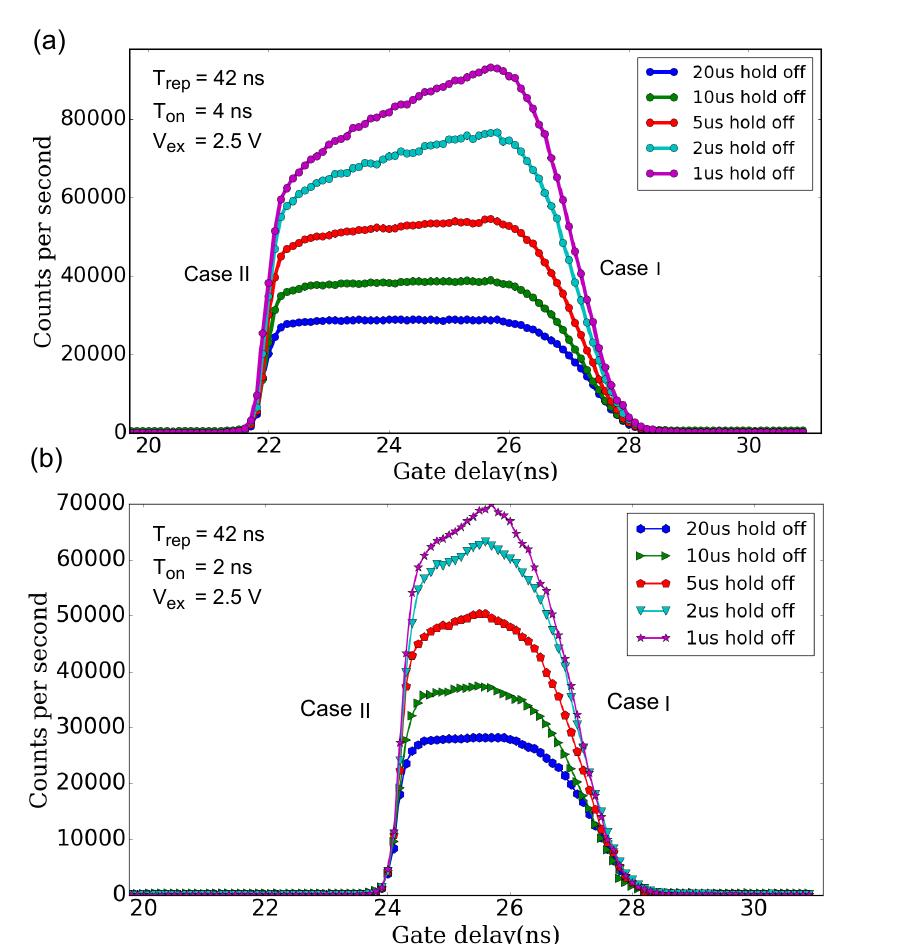}
 	\centering
 	\caption{Effect of hold-off time on photon counts for (a) $T_{\text{on}}$ = 4 ns and (b) $T_{\text{on}}$ = 2 ns. Other SPD parameters are fixed for both the cases, $V_{\text{ex}}$= 2.5 V, temperature= -$40^\circ\text{C}$}
 	\label{fig:Counts_vs_gate delay}
\end{figure}

\section{Results and discussions}

\subsection{Afterpulse probability}
The lifetime of a trapped carrier, also indicative of the decay in afterpulse probability, has been modeled mathematically by various research groups \cite{cova1991trapping,korzh2015afterpulsing,humer2015simple}. 
Using  (\ref{final equation for afterpulse}),  we measured the afterpulse probability for various hold-off times and, observed that it follows a power-law of the form \cite{ziarkash2018comparative, itzler2012power} (with $\Delta T \triangleq t$),
\begin{equation}
 \text{APP}(t) =A_0\,t^{-\lambda}+d,
\label{Decay equation}
\end{equation}
where $\lambda$ is a effective decay constant, $A_0$ is the initial afterpulse probability and $d$ an offset due to background noise counts.

\begin{figure}[tbh]	
 	\centering
 	\includegraphics[width=0.71\columnwidth]{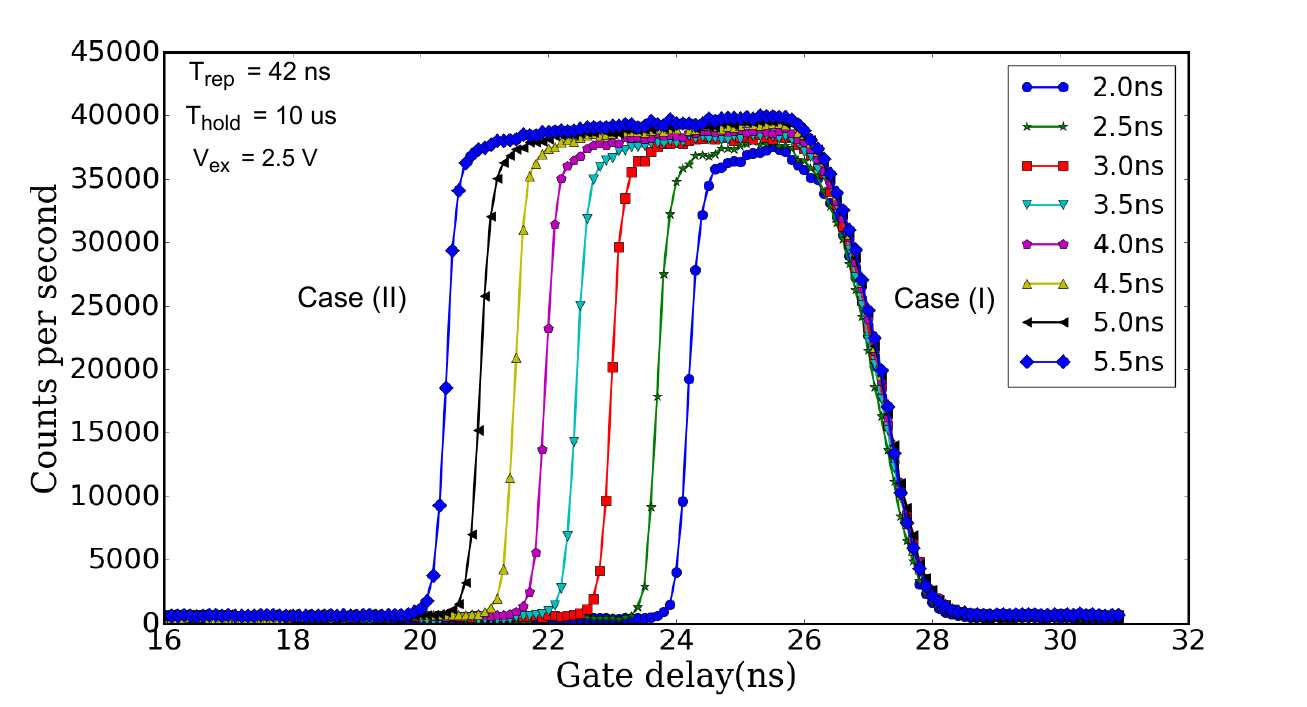}
 	\centering
 	\caption{Effect of gate ON time on number of clicks in SPD. We observe a higher count probability for Case (I) as compared to Case (II).}
 	\label{fig:Photon counts for varying gatewidth}
\end{figure}

Fig.~\ref{fig:Photon counts for varying gatewidth} shows the clicks/s for various gate ON times. 
In our experiments, we fixed the electronic delay at $100\,\text{ps}$, to accommodate the phase noise of the FPGA board clock ($\leq 30\,$ps). We swept the gate pulse, of various widths, in steps of 100~ps through the  MLL pulse repetition time of 42~ns and measured the photon counts per second incident on the SPD. Fig.~\ref{fig:Counts_vs_gate delay} (a) and (b) represent the variation of total counts/s  with respect to hold-off times for a gate width of 4~ns and 2~ns respectively, keeping excess bias voltage $(V_{\text{ex}})$ at 2.5 V and temperature at -$40^\circ\text{C}$. In both Fig.~\ref{fig:Counts_vs_gate delay} (a) and (b), we also observe a higher number of clicks/s near the rising edge of the gate, Case (I), in comparison to the falling edge of the gate, Case (II). This effect is more pronounced as we decrease the gate hold-off time from $20\,\mu\text{s}$ to $1\,\mu\text{s}$, leading us to conclude that its origin lies in the incomplete quenching of charge carriers after an avalanche process.

\begin{figure}[tbh]	
 	\centering
 	\includegraphics[width=0.73\columnwidth]{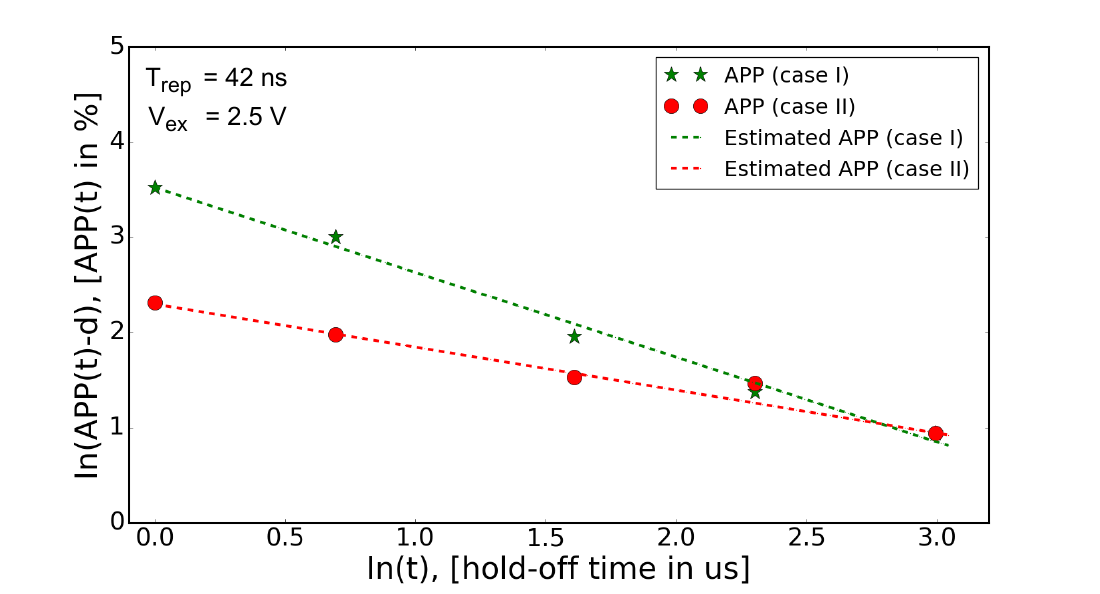}
 	\centering
 	\caption{Rate of afterpulse probability decay is different for both cases (I and II).  Using (\ref{Decay equation}), $A_0$ was estimated to be $33.8\%$ and  $9.92\%$ for case (I) and case (II) respectively.}
 	\label{fig:afterpulse probability for various holdoff}
\end{figure}

Referring to (\ref{Decay equation}), our experimental data fits reasonably well with both Cases (I) and (II), albeit with different decay rates, as seen in Fig.~\ref{fig:afterpulse probability for various holdoff}.  The slope of the linear fits give us $\lambda \sim 0.9$  and 0.4, for Cases (I) and (II), respectively. Similarly, the $y$ intercept gives us the value of  $A_0$, estimated as $33.8\%$ and  $9.92\%$ for Cases (I) and (II) respectively. We found that there is a significant reduction in the value of decay constant, and also a reduction in afterpulse probability,  when a photon arrives closer to the falling edge of a gate than when it arrives just after the rising edge of gate. A power-law fit gives us an afterpulse decay time of $1.1$ to $2.4\,\mu$s.

\begin{figure}[tbh]	
 	\centering
 	\includegraphics[width=0.73\columnwidth]{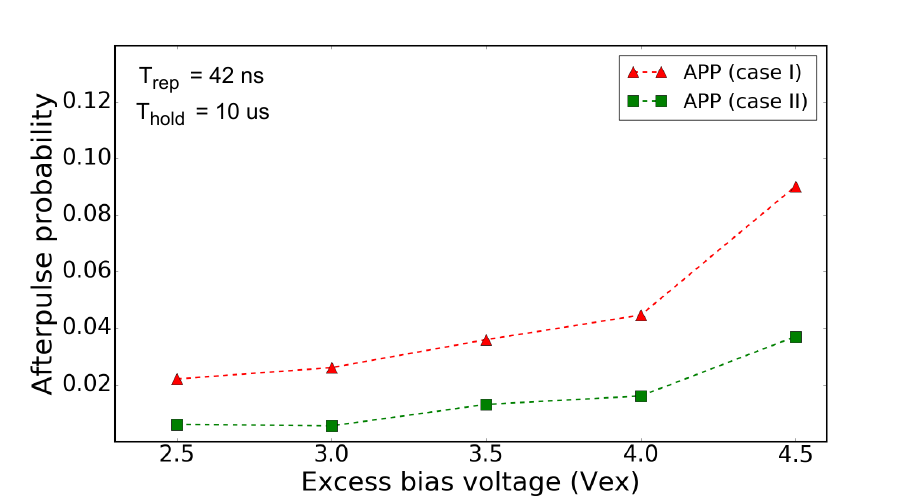}
 	\centering
 	\caption{Effect of excess bias voltage on afterpulsing effect for gate width of 2 ns, while keeping temperature at -$40^\circ\text{C}$, hold-off time is 10 $\mu$s. Afterpulse probability increases with higher excess bias voltages.}
 	\label{fig:APP dependence on Vex}
\end{figure}

\begin{figure}[tbh]	
 	\centering
 	\includegraphics[width=0.73\columnwidth]{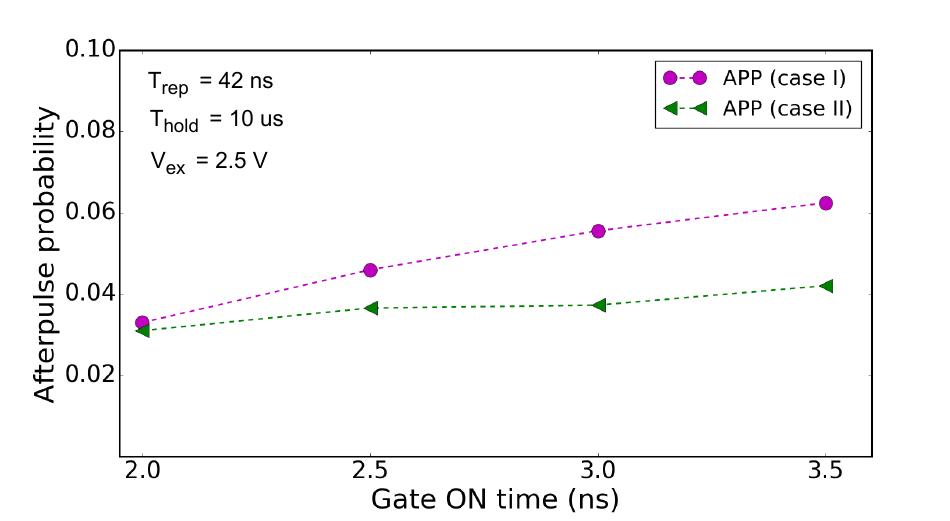}
 	\centering
 	\caption{Afterpulse probability is less than 3.5$\%$ for a gate width of 2 ns.}
 	\label{fig:APP_gatewidth}
\end{figure}

Using the method described in Fig. \ref{method}, we measured afterpulse probability for various excess bias voltages. The dependency of afterpulse probability on  $V_{\text{ex}}$, is shown in  Fig.~\ref{fig:APP dependence on Vex}. 
Following (\ref{final equation for afterpulse}), we also note down the influence of gate window size on the afterpulsing effect. It is  apparent from Fig.~\ref{fig:APP_gatewidth} that the afterpulse effect is more serious with a wider gate ON time. Most of QKD experiments demand a low QBER. Hence, choosing the smallest possible gate ON time keeps the afterpulse contribution as low as possible.

\subsection{Dark count noise}
\begin{figure}[tbh]	
 	\centering
 	\includegraphics[width=0.73\columnwidth]{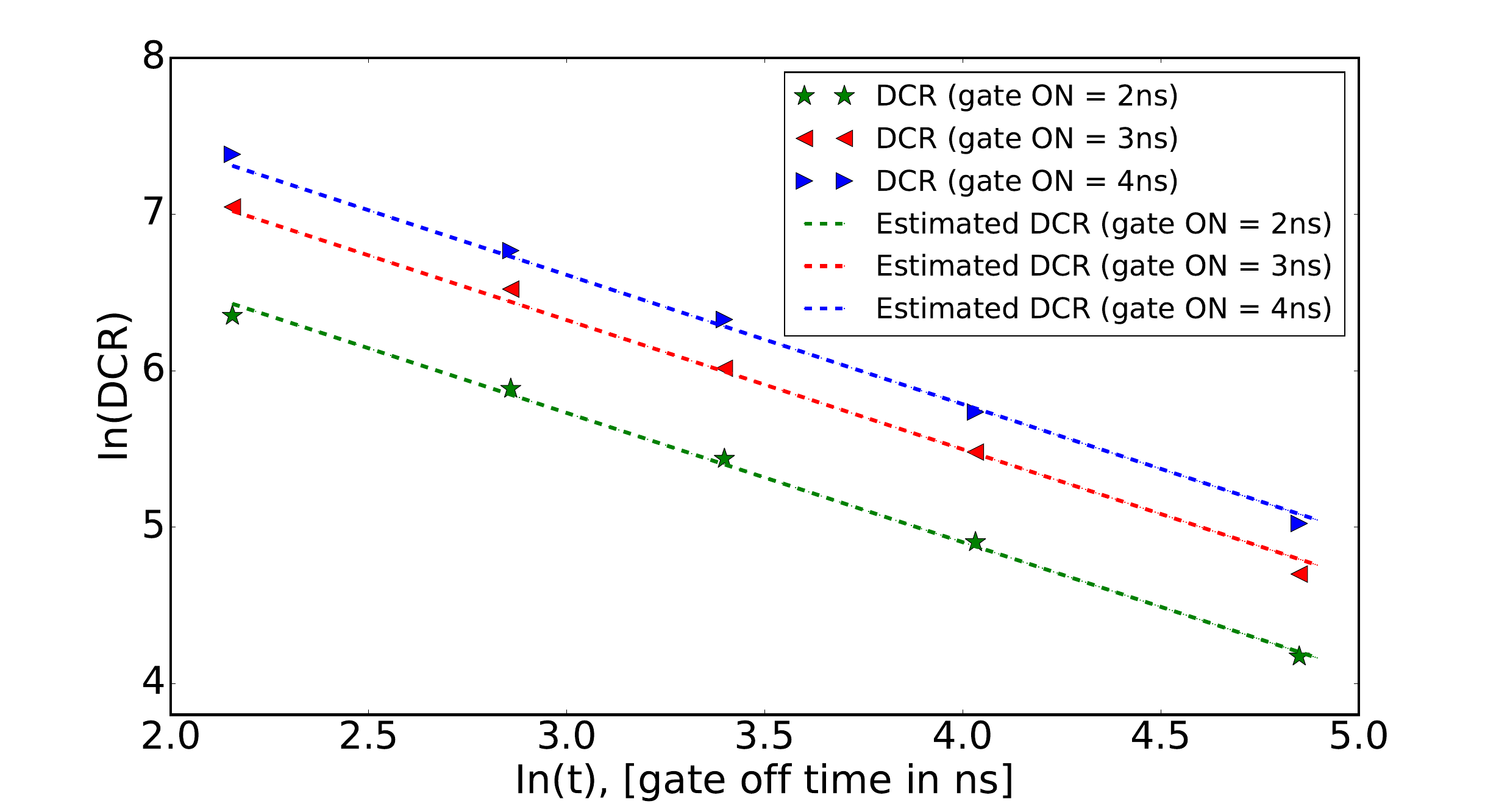}
 	\centering
 	\caption{DCR for various gate ON time with varying gate off time. Using (\ref{Decay equation}),  $D_0$ was estimated to be $10680$, $8003$ and $4420$  for  gate ON time of 4 ns, 3 ns and 2 ns respectively.}
 	\label{fig:DCR_vs_gateoff_gate on}
\end{figure}

To understand the dark count contributions in gated mode SPD under various conditions, we plot the dark count rate (DCR) vs gate off time, on a logarithmic scale, for various gate ON times. \textcolor{black}{Similar to the afterpulse probability decay (\ref{Decay equation}), our measured DCR is well described by three parallel lines (\ref{Decay equation_DCR}) corresponding to gate ON time of 4 ns, 3ns and 2 ns, as shown in Fig.~\ref{fig:DCR_vs_gateoff_gate on}. The decay constant  for DCR was estimated to be  1.03 for various  gate ON times. The linear fit in Fig.~\ref{fig:DCR_vs_gateoff_gate on},  is based on the singular-value decomposition, such that the sum of squares of the distances of the given points to the three lines, with the same slope ($\lambda$) but different intercepts $A_0$, is minimized \cite{gander1995some}.}
\begin{equation}
 \text{DCR} = D_{0}\,t^{-\lambda},
\label{Decay equation_DCR}
\end{equation}
where $\lambda$ is a decay constant, $D_0$ represents initial dark count rate and $t$ here represents gate off time. 
Fig.~\ref{fig:DCR_vs_gateoff_gate on} shows that initial DCR ($D_0$) is reduced by a factor of 2, when the gate ON time is reduced from 4 ns to 2 ns. A lower gate repetition period, or a higher rate, results in more frequent intervals during which the device is primed by the bias voltage and produces a higher DCR.

\begin{figure}[tbh]	
 	\centering
 	\includegraphics[width=0.73\columnwidth]{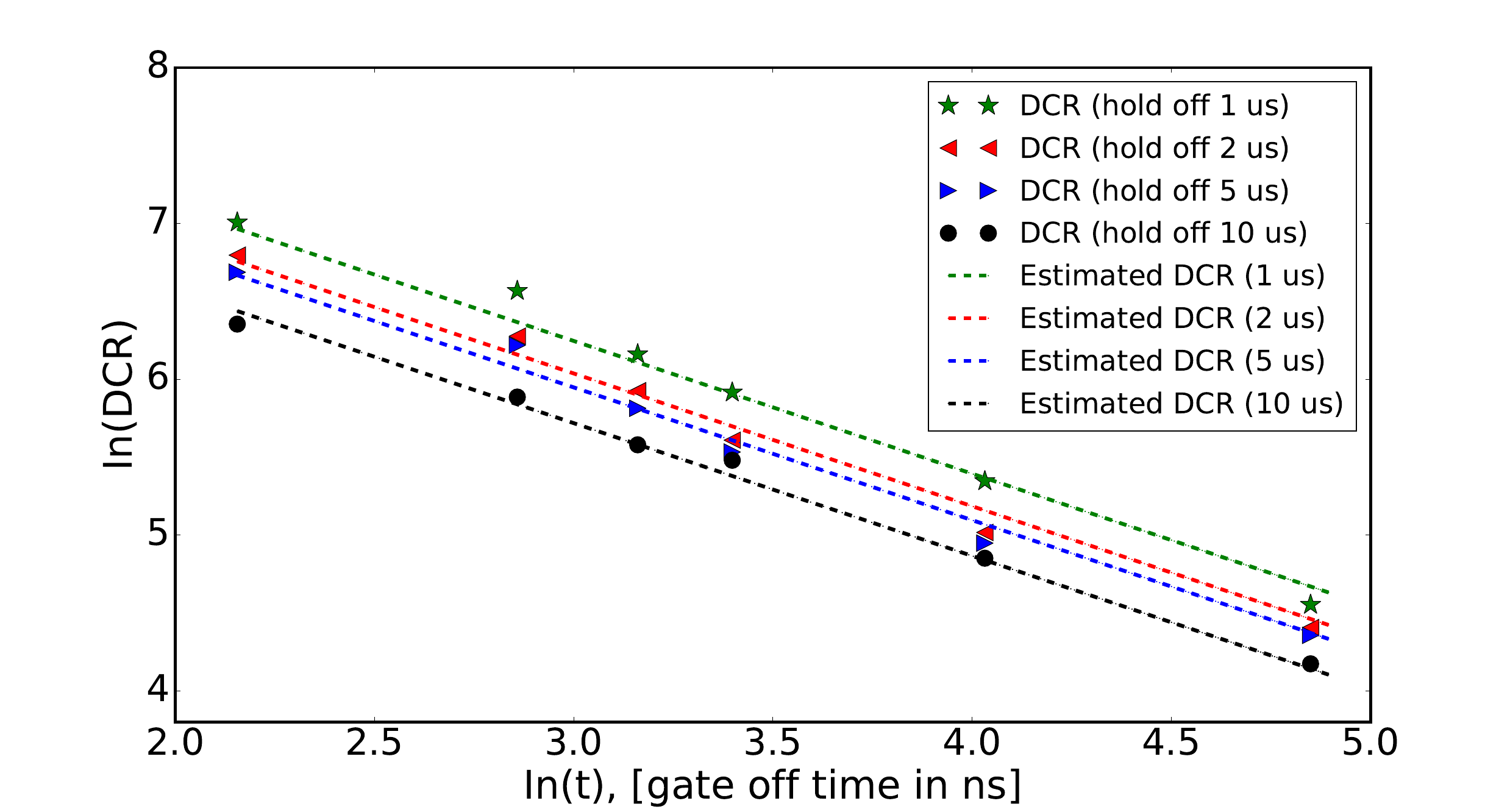}
 	\centering
 	\caption{DCR for various hold-off time with varying gate-off time. Using (\ref{Decay equation}), $D_0$ was estimated to be $10985$, $8490$, $7763$ and $6172$  for  hold-off times of 1$\mu$s, 2$\mu$s, 5$\mu$s and 10 $\mu$s respectively.}
 	\label{fig:DCR_vs_gateoff_holdoff}
\end{figure}

Similarly, dark count contributions for various hold-off times can be analyzed using Fig.~\ref{fig:DCR_vs_gateoff_holdoff}.  We found that initial estimates of DCR are $10985$, $8490$, $7763$ and $6172$ for a  hold-off time of 1 $\mu$s, 2 $\mu$s, 5 $\mu$s and 10 $\mu$s respectively, and the estimated  decay constant of dark carriers is 1.03. At a lower hold-off time, the SPD multiplication region is exposed to an excess bias voltage more frequently and hence it introduces more charge carriers, ultimately causing higher dark counts. Since the gate repetition rate was kept near 24~MHz, with a gate ON time of 2~ns for our afterpulse measurement, the dark count contribution is $\approx$ 0.6$\%$. The DCR contribution reduces by about a factor of 2, when $T_{\text{hold}}$ is increased from 1 $\mu$s to 10 $\mu$s. 

\subsection{Timing jitter}
Timing jitter is usually defined as the total time uncertainty between the absorption of a photon  and the generation of an output electrical pulse. The plots in  Fig.~\ref{fig:Counts_vs_gate delay} (a) and (b) don't capture the statistics of SPD output responses for incident photons. However, by adjusting the timing of the arrival photon, relative to the gate pulse, we were able to collect the distribution of photons within the gate ON time. 

A time to digital converter, the TDC7200 from Texas Instruments,  interfaced with a TIVA Launchpad micro-controller, was used to digitize the timing instants of photon arrival at the SPD, thus yielding  the temporal photon distribution within a gate pulse duration. We achieved a good temporal resolution,  of 55~ps, by monitoring the time-stamps associated with the clicks recorded at the SPD. The TDC start pulse is synchronized to the gate pulse and the SPD generates the TDC stop pulse. 

\begin{figure}[tbh]	
 	\centering
 	\includegraphics[width=0.71\columnwidth]{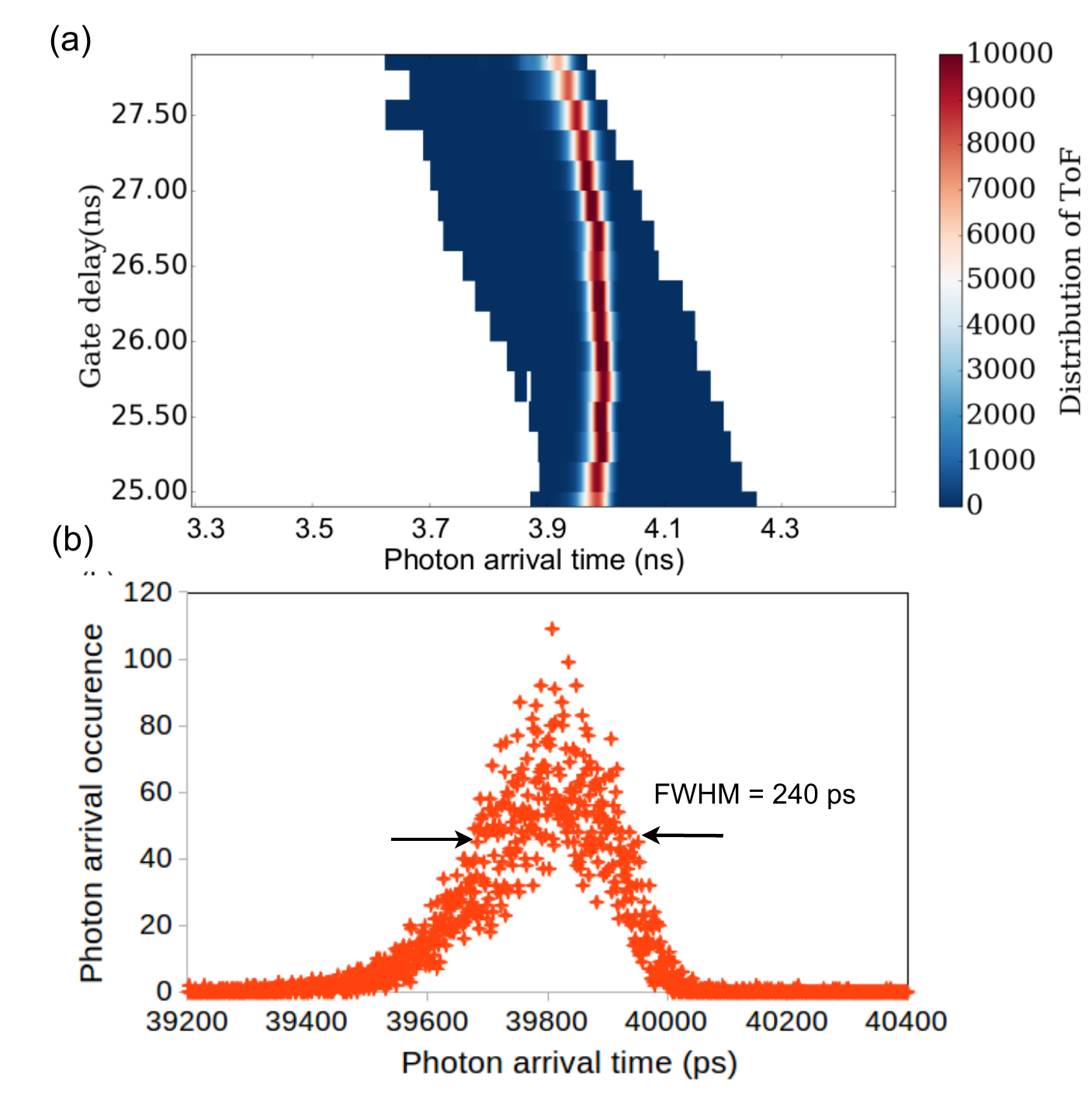}
 	\centering
 	\caption{(a) Distribution of photon detection response within a gate window of 2 ns for various gate delays (b) Timing jitter of about 240~ps is measured as full-width half-maximum (FWHM) of the  response function of the SPD.}
 	\label{fig:SPD jitter}
\end{figure}

Fig.~\ref{fig:SPD jitter}(a) is a surface plot of time of flight (ToF) to the SPD, for incident photons, at various gate delays.
Quantitatively, the timing jitter can be calculated  by measuring the full-width half-maximum
(FWHM) of the ToF statistics at the SPD, as shown in Fig.~\ref{fig:SPD jitter}(b). 

The timing jitter of the SPD, and the source, finally limit the maximum key transmission  rate in a QKD system. e.g. the coherent one-way QKD protocol \cite{stucki2009continuous} uses timing to differentiate between logic bits $0_{\text{L}}=01$ and $1_{\text{L}}=10$, and we can support a raw key transmission of 1~Gbps. 

\section{Conclusion}
We have presented a new approach to estimate afterpulse probability for gated InGaAs single-photon detector with ultrashort weak coherent pulses having a sub-ps pulse width. We found that there is a significant reduction in afterpulse probability  when a photon falls towards falling edge of gate than that of the rising edge of gate. A power law fit gives us an afterpulse decay time of $1.1$ to $2.4\,\mu$s. We found that the SPD performs optimally with $\text{APP} < 3\%$, when the excess bias is set to 2.5~V and a  hold-off time of $10\,\mu$s. A time to digital converter, kept synchronous to the FPGA board, was used to capture the ToF statistics of SPD, from which we estimated a spread of 240~ps in the arrival times.


\section*{Acknowledgment}
We are thankful to Shashank and Dr. Prabha, IIT Madras, for many discussions on QKD.  This work was supported by Ministry of Human Resources and Development (MHRD) vide sanction no. 35-8/2017-TS. GS also acknowledges support on the Visvesvaraya PhD Scheme for Electronics and IT, Govt. of India. 

\section*{References}
\bibliography{afterpulse.bib}
\bibliographystyle{model1-num-names}
\end{document}